\documentclass[showpacs,twocolumn,a4paper]{revtex4}
\usepackage{graphicx}
\usepackage{multirow}

\usepackage{graphicx}% Include figure files
\usepackage{dcolumn}% Align table columns on decimal point
\usepackage{bm}% bold math
\usepackage[colorlinks,linkcolor=blue,citecolor=blue,dvipdfm]{hyperref}
\usepackage{multirow} %multi-rows
\usepackage{booktabs,threeparttable} %threeparttable

\begin{document}

\title{Origin of the Pseudogap and Its Influence on Superconducting State}

\author{Y. Zhou$^1$}
\author{H. Q. Lin$^2$, and C. D. Gong$^3,^1$}
\affiliation{$^1$National Laboratory of Solid State Microstructure, Department of
Physics, Nanjing University, Nanjing 210093, China \\
$^2$Department of Physics and the Institute of Theoretical Physics, Chinese
University of Hong Kong, Hong Kong, China\\
$^3$Center for Statistical and Theoretical Condensed Matter Physics,
Zhejiang Normal University, Jinhua 321004, China}

\begin{abstract}
When holes move in the background of strong antiferromagnetic correlation,
two effects with different spatial scale emerge, leading to a much reduced
hopping integral with an additional phase factor. An effective Hamiltonian
is then proposed to investigate the underdoped cuprates. We argue that the
pseudogap is the consequence of dressed hole moving in the antiferromagnetic
background and has nothing to do with the superconductivity. The momentum
distributions of the gap are qualitatively consistent with the recent ARPES
measurements both in the pseudogap and superconducting state. Two thermal
qualities are further calculated to justify our model. A two-gap scenario is
concluded to describe the relation between the two gaps.
\end{abstract}

\maketitle

\section*{1. Introduction}

One of the most fascinating properties of the cuprates is the opening of the pseudogap above the superconducting (SC) critical temperature $T_{c}$ in the underdoped and optimally doped regime, where most abnormalities observed\cite{TTimusk}. The origin of the pseudogap and its relation to the SC gap are fundamental questions to realize the physics underline the high-temperature superconductors. Although much progress has been made, the issues remain open. There are two distinct scenarios on the relation between the pseudogap and SC gap\cite{STS}. 1) The one-gap scenario: The pseudogap is viewed as the precursor of SC gap, reflecting pair fluctuation above $T_{c}$, and would acquire phase coherence below $T_{c}$.\cite{Emery,PALee} The argument was based on angle-resolved photoemission spectroscopy (ARPES)\cite{HDing,Loeser,Valla}, electron tunneling \cite{Renner,Campuzano}, and thermal transport measurements\cite{Hawthorn}, etc.; 2) The two-gap scenario: The pseudogap is not directly related to the SC gap, but emerges from some ordered states such as antiferromagnetism\cite{AF}, staggered flux\cite{SF}, stripe\cite{STRIPE-Emery,STRIPE-Kivelson}, spin or charge density wave\cite{SCDW-Moon,SCDW-Greco}, orbital circulating currents\cite{OCC}, and resonating valence bond spin liquid\cite{YRZ}, etc., and competes with SC gap. Many experiments, including ARPES\cite{Kaminski,Hashimoto,RHHe}, electronic Raman scattering\cite{LeTacon}, and elastic neutron diffraction\cite{Lipscombe,Fauque}, etc., seem to favor the two-gap scenario. More recently, the improved ARPES data\cite{Tanaka,WSLee,TKondo,KTerashima,Kanigel,Kanigel2} showed that the gap opens near the gap (antinodal) region, and vanishes near the arc (nodal) region in the pseudogap state, which leads to the arc structure of the Fermi surface. In the SC state, the pseudogap dominates the gapped region, while in the arc region, the simple d-wave type gap dominates.

In normal state of conductors, the low energy excitation can be described in terms of quasiparticle picture by introducing the self-consistent single particle scattering potential. Such simple quasiparticle picture failed to describe the conventional superconducting state due to the emergent particle non-conservation induced by the superconducting condensation. However, the quasiparticle picture can be recovered by further introducing the self-consistent pair potential. It is natural to ask whether the quasiparticle picture remains available in cuprates, especially in the most interested pseudogap state. The widely used slave-boson\cite{Brinckmann} and Gutzwiller projection approximation\cite{ZFC} are suitable rather for the overdoped than the underdoped case as compared with experiments. Models account for those above mentioned ordered states capture some aspects of the pseudogap state, but a theoretical framework, which gives a full picture from the pseudogap to SC state, is not yet established. Recently, Yang et al. proposed a phenomenological model\cite{YRZ} to account for the underdoped cuprates. However, both the pseudogap and SC gap adopted there are experimentally fitted for different doping in reality. We believe that a reasonable quasiparticle picture can be also proposed in the pseudogap region if the main strong correlation effects have been taken into account. This is significantly important even if it is approximate to some degrees. The questions are now how to extract the main effects and to what degree the quasiparticle picture consists with the experimental measurements.

In this paper, we try to establish an effective single particle Hamiltonian to understand the underdoped cuprates. The strong correlations in the underdoped cuprates are taken into account. The limitation of no double occupancy leads to much reduced effective hopping integral, reflecting the effect with large spatial scale, and the antiferromagnetic (AFM) background introduces an additional phase factor in the hopping term, reflecting the effect with small spatial scale. By considering these two effects with different spatial scale, the doping and momentum dependence of the pseudogap, arc structure Fermi surface, and the evolution of the quasiparticle dispersion are obtained and they are qualitatively consistent with the experimental discovery. We argue the pseudogap, originated from the short-range AFM correlations, is the property of the single particle and irrelevant to the superconductivity. The model Hamiltonian is extended into the superconducting state by adopting a phenomenological SC term. The results are also qualitatively consistent with recent ARPES measurements. The calculations on the specific heat and superfluid density confirm the validity of proposed effective Hamiltonian. Since the most previous studies on the pseudogap and SC gap focus in the momentum space both experimentally and theoretically, we also restrict our investigations in the momentum space to compare with previous experimental and theoretical results. The possible spatial inhomogeneity below $T^{\star}$, which had been discovered experimentally\cite{Sonier,Parker}, will not be taken into account.

The paper is organized as follows. In Sec. \hyperref[S2]{2}, two main strong correlation effects in underdoped cuprates are considered. The effective single-particle Hamiltonian is consequently proposed. This Hamiltonian is discussed in the pseudogap state, and SC state in Sec. \hyperref[S3]{3}, and Sec. \hyperref[S4]{4}, respectively. In Sec. \hyperref[S5]{5}, We calculate the specific heat and superfluid density to further justify our effective Hamiltonian. The conclusion is given in Sec. \hyperref[S6]{6}, together with some discussions.

\section*{2. Two Spatial Scale Effects and effective single particle Hamiltonian}\label{S2}

The strong correlations play the key roles in the underdoped cuprates. Essentially, this includes two main aspects. One is the AFM background, and the other is limitation of the double occupancy. A sophisticated theory should take into account both of them. We start from the extended $t-J$ model, which is thought to capture the main aspects of the cuprates
\begin{equation}
H=-\left( t\sum_{\left\langle ij\right\rangle _{1}\sigma }+t^{\prime
}\sum_{\left\langle ij\right\rangle _{2}\sigma }+t^{\prime \prime
}\sum_{\left\langle ij\right\rangle _{3}\sigma }\right) c_{i\sigma
}^{+}c_{j\sigma }+J\sum_{\left\langle ij\right\rangle _{1}}\vec{S}_{i}\cdot
\vec{S}_{j},
\end{equation}%
where $\left\langle ij\right\rangle _{n}$ with $n=1$, $2$, and $3$ denoting
the nearest neighbor (NN), second-NN, and third-NN, respectively.

\begin{figure}[tbp]
\vspace{0.0in} \hspace{-0.0in}
\center
\includegraphics
[width=3.5in]{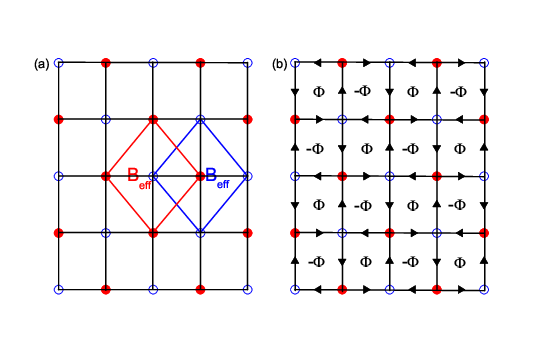} \vspace{-0.0in} \vspace{-0.0in}
\caption{(Color online) (a) Schematic effective magnetic field as described in the
text, and (b) staggered flux phase. The solid and blank circles are for
the atoms of cuprate. It is divided into two sublattice due to the
broken of translational symmetry.}\label{f.1}
\end{figure}

The long-range AFM order disappears under slightly hole doping. However, the neutron scattering\cite{Lipscombe,Fauque} and quantum oscillations measurements%
\cite{Sebastian} show that the AFM correlation and its associated staggered flux state remains finite on the $CuO_{2}
$ plane in the underdoped regime. Their existence has also been found
numerically\cite{Leung1,Ivanov,Leung2}. This gives the substantial contributions to the
underdoped cuprates. Unfortunately, Such effect is usually neglected in the previous
treatments.
In the Table \ref{T.1}, the NN spin-spin correlation $\langle S_{i}\cdot
S_{j}\rangle $ is shown by applying the exact diagonalization (ED) technique
on the $20$-site lattice. In the underdoped cases, it is always negative,
implying the AFM correlation. In the background of the antiferromagnetism,
for each site, its four nearest neighbors compose a plaquette. When a hole
is introduced into the center of this plaquette, it will sense an effective
magnetic field $B_{eff}$ originated from its four neighboring opposite spins
as shown in the Fig. \ref{f.1}(a). $B_{eff}$ can be phenomenologically
estimated by the NN spin-spin correlation between a spin $1/2$ sitting on
the site of the given hole and its surrounding spins%
\begin{equation}
B_{eff}=J\sum_{j\in NN}\langle S_{i}\cdot S_{j}\rangle /g\mu _{B}\langle
S_{z}\rangle
\end{equation}%
with $\langle S_{z}\rangle =1/2$, and $g$, $\mu _{B}$ representing Land\`{e}
factor, Bohr magneton, respectively. This effective magnetic field alters
its sign when the given site moves to the nearest neighbor. Correspondingly,
there is a doping dependent gauge flux $\Phi =B_{eff}a^{2}$ threading through
the plaquette ($a=0.40nm$ denoting the lattice constant) as shown in Fig. \ref%
{f.1}(b). In its journey of wandering, the hole is affected by the
staggered flux field in which $\Phi $ and $-\Phi $ appear alternatively.
This staggered flux phase can be also established by consideration of the
current-current correlation as manifested in the variational Monte
Carlo\cite{Ivanov} and ED calculations\cite{Leung2}. Therefore, in the small
spatial scale, when the hole hops, it will be exerted by a phase shift $\delta /4$
(or $-\delta /4$ ) due to Aharonov-Bohm effect, where $\delta =2\pi \Phi /\Phi
_{0}$ ($\Phi _{0}=hc/e$ is the flux quanta)\cite{Harris}. This also implies
that the original translational symmetry is broken, consisting with the recent
STM experiments\cite{STM}.

\begin{table}
\caption{The ED results of the spin-spin correlation function and hopping
matrix in the 20-site cluster with different hole concentration. $t=1$, $%
t^{\prime }=-0.3$, $t^{\prime \prime }=0.2$, and $J=0.4$.}
\label{T.1}\vspace{-0.0in} \vspace{-0.0in} \tabcolsep=4pt
\begin{center}
\begin{tabular}{lllll}
\hline\hline
\multicolumn{1}{c}{$n_{h}$} & \multicolumn{1}{c}{$\langle S_{i}\cdot S_{j}\rangle$} & \multicolumn{1}{c}{$I_{1}$} & \multicolumn{1}{c}{$I_{2}$} & \multicolumn{1}{c}{$I_{3}$}\\
\hline
$0$ & $-0.3454$ & $0$      & $0$       & $0$  \\
$1$ & $-0.2745$ & $0.0703$ & $-0.0297$ & $0.0125$  \\
$2$ & $-0.2179$ & $0.1483$ & $-0.0350$ & $0.0355$  \\
$3$ & $-0.1670$ & $0.2116$ & $-0.0469$ & $0.0510$  \\
$4$ & $-0.1399$ & $0.2635$ & $-0.0594$ & $0.0732$  \\
$5$ & $-0.0855$ & $0.3126$ & $-0.0709$ & $0.0960$  \\
\hline\hline
\end{tabular}
\end{center}
\end{table}

On the other hand, it is well known that the effective hopping is
much reduced in the cuprates due to the limitation of no double occupancy.
To account for this effect,
we directly calculate the hopping matrix, $I_{n}=\left\langle
C_{i}^{+}C_{j}+C_{j}^{+}C_{i}\right\rangle $ ($n=1$, $2$, $3$ for
the NN, second-NN, and third-NN, respectively), by ED technique as shown in Table. \ref{T.1}. The
hermiticity is guaranteed under such selection. Since its value is
$1-x^{2}$ if the strong correlations are not taken into
account\cite{ZFC}. For low doping, we neglect the high order term
$x^{2}$. Therefore, $I_{n}$ obtained here is just the effective
hopping when $t$ is taken as unit. The average process takes
account for the all possible configurations. In this sense, it is the
effect with large spatial scale. In fact, $I_{n}$ obtained here is similar to the previous renormalized factor adopted in the Slave-boson and Gutzwiller projection approximations\cite{Brinckmann,ZFC}. We find that this value is underestimated in the slave-boson treatments, and
overestimated in the Gutzwiller projection treatments for the NN
hopping. For the second-NN hopping, it approaches to the value under
the slave-boson treatment. Here, we would like to point out that the
NN hopping matrix enhances with decreasing $J$. When $J=0$, its
value is almost the same as that under the Gutzwiller approximation,
especially for $t'=0$, and $t''=0$. Present treatment is a full
consideration of the strong correlation, both the no-double
occupancy restriction and the antiferromagnetic correlation are
taken into account, compared with only the former is considered in
the Gutzwiller projection approximation. For more hole cases, the
hopping matrix should be further divided by a factor approximately
equal to hole number.

The original lattice is now divided into two sublattices $D$ and $E$
with the corresponding fermionic operators $d$ and $e$ due to the
broken translational symmetry as mentioned above\cite{YQS}. The
effective hopping is much reduced due to the effect with large spatial
scale. Furthermore, an additional phase shift is introduced when the
hole hops to the nearest neighbor site due to the effect with the
small spatial scale. Consequently, we may write the following effective single
particle Hamiltonian as,
\begin{equation}
H=\sum_{k\sigma }\left( \gamma _{k}d_{k\sigma }^{+}e_{k\sigma }+hc\right)
+\sum_{k\sigma }\epsilon _{k}\left( d_{k\sigma }^{+}d_{k\sigma }+e_{k\sigma
}^{+}e_{k\sigma }\right),\label{E1}
\end{equation}%
where $\gamma _{k}=-2(I_{1}+J^{\prime }\kappa _{0})(e^{i\delta
/4}\cos k_{x}+e^{-i\delta /4}\cos k_{y})$, and $\epsilon
_{k}=-4I_{2}\cos k_{x}\cos k_{y}-2I_{3}\left( \cos 2k_{x}+\cos
2k_{y}\right) -\mu $, with $\mu $, a chemical potential determined
by the hole density. $\kappa _{0}=\sum_{\sigma }\left\langle
c_{i\sigma }^{+}c_{j\sigma }\right\rangle /x$ is the uniform bond
order with $J^{\prime }=3J/8$ as usual\cite{YRZ}. The summation is
restricted in the magnetic Brillouin zone (MBZ). Numerically, $%
t $ (about $0.4eV$ in the real material) is taken as energy unit, $t^{\prime
}=-0.3$, $t^{\prime \prime }=0.2$, and $J=0.4$ is adopted. As shown in the Table~\ref{T.1},
the hopping matrix $I_{3}$ is strong than $I_{2}$ for more hole doping due to the size effect,
$I_{3}$ is then modified to be $2I_{2}/3$.

So far, an effective model Hamiltonian is established based on our knowledge of the moving hole in the background of antiferromagnetism. We emphasize that no special parameters have been introduced in the discussion of normal state. The concept of staggered flux in cuprates is not a new concept. However, it is rather phenomenological and qualitative previously. Here, we explicitly demonstrate the origin of staggered flux and give the way to determine it quantitatively.

\section*{3. Results in the Pseudogap State}\label{S3}

\begin{figure}[tbp]
\vspace{0.0in} \hspace{-0.0in}
\center
\includegraphics
[width=3.5in]{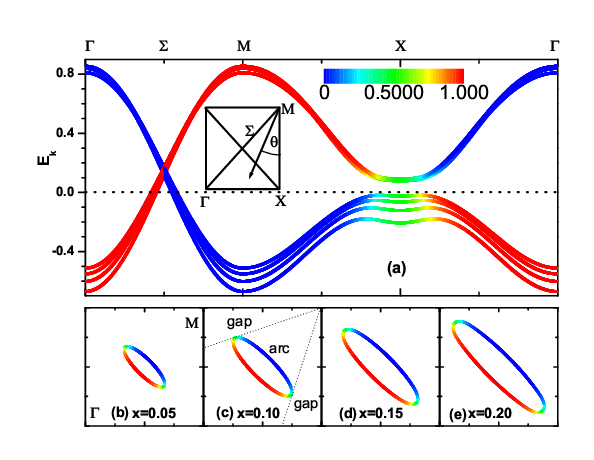} \vspace{-0.0in} \vspace{-0.0in}
\caption{(Color online) Upper panel: the QP dispersion $\epsilon _{k}^{\pm
}$ for different hole concentration (from bottom to top at $X$
point in the lower band are $x=0.05$, $0.10$, $0.15$, and $0.20$.),
together with its weight factor. Thick, and thin lines are for lower ($-$) and upper ($+$) bands, respectively. The Fermi energy is fixed at $0$.
Lower panels: the doping dependent evolution of the Fermi surface.
The dotted lines divide the first quadrant of Brillouin zone into
gap region and arc region.} \label{f.2}
\end{figure}

In the following discussion, we would explore the adaptation of the established effective single particle Hamiltonian (\ref{E1}) to describe the energy spectrum in strongly correlated systems. The effective Hamiltonian is diagonalized with the quasiparticle (QP)
dispersion $\epsilon _{k}^{\eta }=\epsilon _{k}+\eta |\gamma _{k}|$ ($\eta
=\pm$, $+$ for upper band, and $-$ for lower band). Correspondingly, the
diagonal and off diagonal Green's function is given as $G_{k,\omega
}=\sum_{\eta }W^{\eta }/(\omega -\epsilon _{k}^{\eta })$, and $\tilde{G}%
_{k,\omega }=\sum_{\eta }\tilde{W}^{\eta }/(\omega -\epsilon _{k}^{\eta })$
respectively. The weight factor $W^{\eta }=\frac{1}{2}(1+\eta cos\varphi
_{k})$, and $\tilde{W}^{\eta }=-\eta \frac{i}{2}sin\varphi _{k}$ with $%
e^{i\varphi _{k}}=\gamma _{k}/|\gamma _{k}|$. The QP dispersion
together with its weight is shown in upper panel of Fig. \ref{f.2}.
The upper and lower band coincide at $(\pi /2,\pi /2)$. This means
that no full gap opens, unlike the Mott insulator state. The lower
band is rather rigid against doping along the nodal line with much
reduced bandwidth $0.65-0.75t(\sim 260-300meV)$. In the antinodal regime, clear flatness below Fermi energy $%
E_{f}$ (fixed at $0$) up to $x=0.25$ can be found. Therefore, a pseudogap
(denoted by $\Delta ^{PG}$) opens around this regime. The distance between
the flatness part and $E_{f}$ decreases with increasing doping. These
features qualitatively coincide with the ARPES data\cite{Marshall}. We also
plot the evolution of the Fermi surface (FS) in the lower panel of Fig. \ref{f.2}. The intensity is strong near the nodal region, and vanishes
in the antinode region due to the opening of the pseudogap, producing a hole pocket structure. On the
other hand, the weight factor $W^{+}$ is much reduced beyond the
MBZ, especially, near the nodal direction. Therefore, the hole pocket looks more like an arc\cite{Norman}. It extends to a large hole FS above
$x=0.25$, where the pseudogap vanishes gradually. Hence, the arc
structure FS is a direct consequence of the opening of pseudogap.
The evolution of the arc structure and its length also qualitatively
agree with the recent ARPES experiment\cite{KMShen} but with a
slightly expanded doping range because of the overestimating of the
spin-spin correlation due to the size effect.

\begin{figure}[tbp]
\vspace{0.0in} \hspace{-0.0in}
\center
\includegraphics
[width=3.5in]{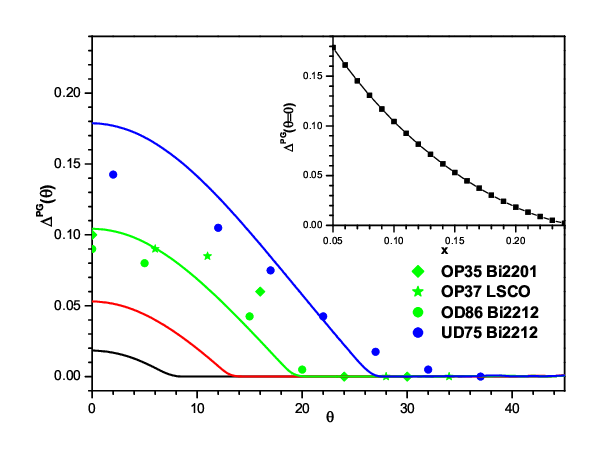} \vspace{-0.0in} \vspace{-0.0in} \caption{(Color online) The
momentum distribution of the pseudogap. The solid line, from top to bottom, is $x=0.05$, $x=0.10$, $x=0.15$, and $x=0.20$, respectively.
The symbols are the experimental data of $Bi2212$ (Ref.~\hyperref[R1]{26}), $Bi2201$ (Ref.~\hyperref[R2]{27}), and $LSCO$ (Ref.~\hyperref[R3]{28}). The experimental data are scaled with $t\sim0.4eV$. Insert is the doping dependence of pseudogap obtained by the linear interpolation of effective parameters.} \label{f.3}
\end{figure}

The doping and momentum evolution of the pseudogap are shown in Fig.~\ref%
{f.3}. The magnitude of pseudogap is determined by evaluating the
minimal distance of the lower band from Fermi energy in the given
direction. $\Delta ^{PG}(\theta )$ decreases with increasing $\theta
$, and disappears at a critical value of $\theta _{c}$. Such
behavior had been confirmed by various
experiments\cite{WSLee,TKondo,KTerashima}. Both the pseudogap region
and its
magnitude decrease with doping as evidenced by ARPES data\cite{Valla}. $%
\Delta ^{PG}(\theta =0)$ decreases from about $72meV$ at deep
underdoped region to about $24meV$ at optimal doping $(x=0.15)$
under the lowest ordered approximation as shown in the insert of
Fig.~\ref{f.3}. In the present treatments, we do not adopt any
pairing potential, which is distinct from the previous slave-boson
calculation\cite{LJX}. Furthermore, the
behavior of pseudogap differs from the early ARPES discovery\cite%
{HDing,Loeser}, and is far from that of the simple d-wave gap. Therefore,
the pseudogap is more likely AFM correlation originated, and seem to be not
related to the SC pairing. This is well supported by the scaling of $T^{\ast
}$ (characterizing of the pseudogap) with the Neel temperature $T_{N}$\cite%
{Keren}. We notice that the magnitude of gap below and above $E_{f}$ near the
antinode is almost the same in the intermediate doping ($x=0.1\sim 0.15$).
That means nearly symmetric gap opens, being analogous to the BCS gap.
However, it should be emphasized that this symmetry is broken beyond both the antinodal
region and the intermediate doping range as shown in Fig. ~\ref{f.2}. Therefore, it does not imply the
preformation of SC pairing, differs from the conclusion of Yang \textit{et al}.\cite{Yang}. The particle-hole asymmetry and possible back-bending phenomena in hole-doped cuprates\cite{Hashimoto, RHHe} can be further performed, which will be discussed elsewhere. This asymmetry is also obtained within the resonating valence bond spin liquid\cite{LeBlanc} or charge density wave model\cite{Greco}. However, the parameters they selected is rather phenomenological.

\section*{4. Results in the Superconducting State}\label{S4}

To extend our model into the SC state, a phenomenological SC term $%
-\sum_{k}\Delta _{k}^{SC}\left( d_{k\uparrow }e_{-k\downarrow }+e_{k\uparrow
}d_{-k\downarrow }+hc\right) $ is further assumed though the essential
mechanism is still unclear now. Here, the SC gap function $\Delta _{k}^{SC}=V\Delta ^{SC}\left( \cos k_{x}-\cos
k_{y}\right) $ follows the standard d-wave symmetry as evidenced by the ARPES measurements\cite{TKondo,KTerashima,Ma,Wei}. $V$ is
the pairing potential and $\Delta^{SC}$ denotes the gap parameter which will be determined self-consistently. Clearly, a two-gap scenario is adopted in the SC state. The total gap $\Delta _{k}^{T}$ contains two components: SC gap $\Delta _{k}^{SC}$ and
pseudogap $\Delta _{k}^{PG}$. Numerically, the magnitude of $\Delta _{k}^{T}$ is obtained in the same way as $\Delta ^{PG}(\theta )$ as discussed in the previous section.

\begin{figure}[tbp]
\vspace{0.0in} \hspace{-0.0in}
\center
\includegraphics
[width=3.5in]{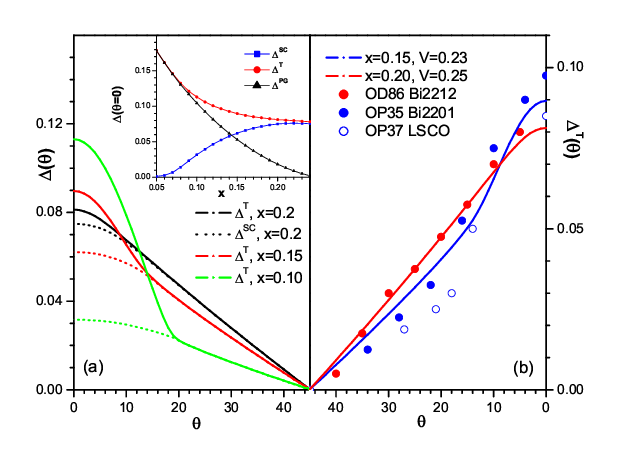} \vspace{-0.0in} \vspace{-0.0in}
\caption{(Color online) (a) The momentum distribution of the total gap in the SC state, together with its corresponding
d-wave SC gap with pairing potential $V=0.23$. Insert shows the doping
dependence of the total gap $\Delta^{T}$, SC gap $\Delta^{SC}$, and
pseudogap $\Delta^{PG}$ at $\theta=0$, the corresponding parameters are obtained by linear interpolation. (b) The comparison of total gap $\Delta^{T}(\theta)$ between the theoretical results and experimental data. Red circles are for $Bi2212$ (Ref.~\hyperref[R1]{26}),
Solid and hollow blue circles are for $Bi2201$ (Ref.~\hyperref[R2]{27}), and $LSCO$
(Ref.~\hyperref[R3]{28}), respectively. All experimental data are scaled with $t\sim0.4eV$} \label{f.4}
\end{figure}

As can be seen from the insert of Fig.~\ref{f.4}(a), the magnitude of $\Delta ^{SC}(\theta )$ (also SC critical
temperature $T_{c}$) increases with increasing doping from $x=0.06$ up to $%
x=0.25$, where the pseudogap vanishes. It should be pointed out that the
magnitude of $\Delta ^{SC}(\theta =0)$ is larger than $\Delta ^{PG}(\theta
=0)$ at about $x=0.15$ at given pairing potential $V=0.23$. These
qualitatively agree with the experimental phase diagram. The conventional one-gap slave boson treatment cannot give such
doping dependence unless the bosonic condensation is assumed\cite{Emery}.
The suppression of the SC gap is a natural product of the opening of
pseudogap. The density of state (DOS) near $E_{f}$ is much reduced as
comparing with the normal state. Additionally, this reduction mainly comes
from the antinodal region. As well known that they are the main factors to
determine the magnitude of d-wave pairing at given pairing potential $V$.

Now, we turn to discuss the momentum distribution of the gap. In the
nodal region, $\Delta ^{T}(\theta )$ increases almost linearly with
decreasing angle $\theta $ as shown in Fig.~\ref{f.4}(a). This is
the typical d-wave behavior, showing that the d-wave SC pairing
dominates this region. In fact, the gap comes from the SC
condensation entirely because of the vanished pseudogap along the
arc. In the antinodal region, the deviation from the standard d-wave
can be found, which weakens upon doping. The shape and value of
$\Delta ^{T}(\theta =0)$ with small value of $\Delta ^{SC}$ are
quite similar to that of the pseudogap state, consisting with the so-called 'soft gap' nature
of pseudogap\cite{Ma}. Therefore, the pseudogap dominates the
antinodal region. Between the two regions, the pseudogap and the SC
gap coexist and compete with each other, so the deviation from the
simple d-wave enhances gradually when approaching to the
antinodal region. The degree of the deviation depends on the ratio of $%
\Delta ^{SC}\left( \theta =0\right) /\Delta ^{PG}(\theta =0)$. Consequently,
the deviation weakens upon doping. Such effect can also explain the
different deviation in optimal doped $Bi2201$\cite{TKondo} and $Bi2212$\cite%
{JMesot}. These features consist with recent experiments\cite%
{WSLee,KTerashima}. The deviation had also been obtained in the previous
one-gap scenario by applying the spin fluctuation theory\cite{LJX}. However,
the correction is too small to account for the large deviation in
underdoped cuprates. We further compare the experimental data with our
theoretical results in Fig.~\ref{f.4} (b). For $Bi2201$%
\cite{TKondo,Ma} and $LSCO$\cite{KTerashima} materials, a reduced pairing
potential $V=0.18$ is adopted. The pseudogap in the two cuprates is almost
same as $Bi2212$\cite{WSLee} at optimal doping, but the SC critical
temperature in the former is much weakened than that in the latter. The
comparison shows qualitative agreement.

\section*{5. Application on Thermodynamic quantities}\label{S5}

\begin{figure}[tbp]
\vspace{-0.1in} \hspace{-0.0in}
\center
\includegraphics
[width=3.5in]{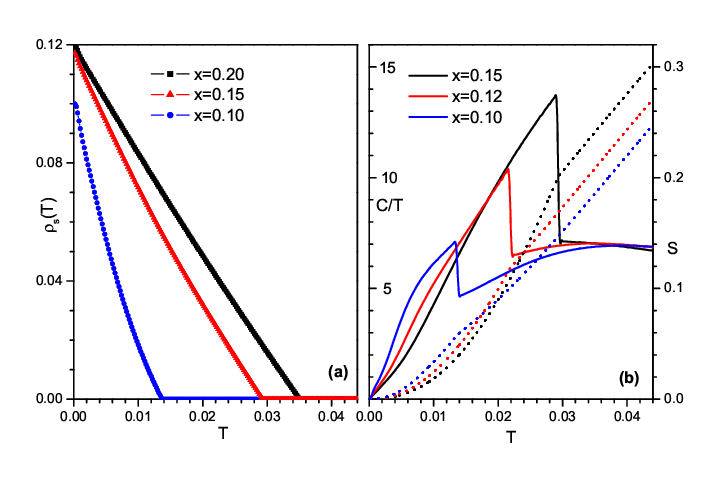} \vspace{-0.0in} \vspace{-0.0in}
\caption{(Color online) (a)Temperature evolution of the superfluid density $\rho%
_{s} $. (b) Temperature dependence of the specific heat coefficient
$C/T$ (left scale) and entropy $S$ (right scale).} \label{f.6}
\end{figure}

To justify our theoretical model, the superfluid density $\rho _{s}$
and specific heat $C/T$ are calculated. As well known that $\rho
_{s}$ shows almost a low-temperature linear $T$ behavior as shown in Fig.~\ref{f.6}(a), expected for
clean BCS d-wave superconductor, i.e., $\rho _{s}\left( T\right) $
$=\rho _{s}(0)-bT$. In the underdoped region, $\rho _{s}(0)$ strongly
depends on the doping $x$, while the coefficient $b$ is less
sensitive to $x$. This can not be understood within a simple BCS
d-wave model. On the other hand, it seems that there is a
contradiction between the large gap and small superfluid density in
the one-gap scenario\cite{Broun,Hetel}. In fact, these curves are slightly concave
upward. This upward weakens with doping, and disappears when the
pseudogap vanishes in the case of heavy overdoping. The magnitude of
$\rho _{s}(0)$ increases with doping in the underdoped region. Our results
are similar with that of Ref.~\hyperref[R4]{55}, in which both
the pseudogap and SC gap are fitted from the experimental data\cite{Nicol}. The results of the specific heat $C/T$ are shown in
Fig.~\ref{f.6}(b). The jump at critical temperature $T_{c}$, i.e.,
$\delta C/T$ increases rapidly upon doping, while the entropy $S$
seem to extrapolates to negative value at $T=0$. These main
features are consistent with the measurements of Loram \textit{et al}\cite%
{Loram}.

\section*{6. Conclusion and discussion}\label{S6}

In conclusion, we have proposed an effective single particle Hamiltonian to
investigate pseudogap in the high-$T_{c}$ cuprates. The dynamic parameters in the Hamiltonian are extracted from the exact diagonalization studies of the $t-J$ model and the consideration of the effective A-B effect. The main idea is to take the effects of two spatial scales into account, one leads to much reduced hopping and the other
adds to a phase factor, both reflect physics of the doped holes moving in AFM
background. Our results revealed that the pseudogap is the
single particle property, and is not directly related to the SC state. It
opens near the antinodal region and vanishes around nodal region. In the SC
state, the simple d-wave SC gap dominates the nodal region, while the
pseudogap dominates the antinodal region. The doping and momentum dependence
of the gap are qualitatively consistent with recent ARPES
measurements both in the pseudogap state and superconducting state. A two-gap scenario is therefore concluded. The superfluid
density and specific heat are further calculated based on the present model,
which consists with the experiments qualitatively. Our simple model seems to
capture the main features of the underdoped cuprates.

The quasiparticle dispersion obtained from our effective Hamiltonian slightly differs from those obtained by the numerical calculations\cite{Nodegap,Sakai}, where the nodal gap above Fermi surface is found. Since only two main effects with different spatial scale are taken into account in present treatment, such deviation is not surprising. However, we would like to point out that the momentum- and doping dependence of the pseudogap and gap in the superconducting state are insensitive to the nodal gap due to the fact that it is well above the Fermi level (about $40\sim 60mev$). On the other hand, as shown in the Fig.~\ref{f.2}, the nodal arc length is slightly larger than the experimental results\cite{KMShen}. This can be modified by further considering the random phase approximations. However, we would like to emphasize that this modification does not change the main tendency of the doping dependence and momentum distribution of the gap. In the present
treatment, only the influence of pseudogap on the SC gap is considered. The feedback effect of the SC gap to the pseudogap is not considered, which
is expected to suppress the spin-spin correlation and will be studied in future.

%\begin{acknowledgment}
Y Zhou acknowledges helpful discussions with A Greco , and JX Li. This
work was supported by NSFC Projects No. 10804047, 11274276, and A Project Funded by the Priority
Academic Program Development of Jiangsu Higher Education
Institutions. CD Gong acknowledges 973 Projects No. 2011CB605906. HQ Lin acknowledges RGC Grant of HKSAR, Project No. HKUST3/CRF/09.
%\end{acknowledgment}

\end{document}